# Decoupling of Two Closely Located Dipoles by a Single Passive Scatterer for Ultra-High Field MRI

M.S.M. Mollaei[*,1], S.A. Kurdjumov[2], A.A. Hurshkainen[2] and C.R. Simovski[1]

**Abstract** – We report decoupling of two closely located resonant dipole antennas dedicated for ultra-high field magnetic resonance imaging (MRI). We show that a scatterer slightly raised over the plane of antennas grants a sufficient decoupling even for antennas separated by very small gap (below 1/30 of the wavelength). We compare the operation of two decoupling scatterers. One of them is a shortcut resonant dipole and another is a split-loop resonator (SLR). Previously, we have shown that the SLR offers a wider operational band than the dipole and the same level of decoupling. However, it was so for an array in free space. The presence of the body phantom drastically changes the decoupling conditions. Moreover, the requirement to minimize the parasitic scattering from the decoupling element into the body makes the decoupling dipole much more advantageous compared to the SLR.

## 1. INTRODUCTION

Decoupling of transceiver antennas used in ultra-high field magnetic resonance imaging (MRI) is a crucial and challenging task [1-4]. The further development of 7 T MRI for prostate and brain requires 32 transceiver dipoles array [5-7] whereas the Larmor's frequency and, consequently, the carrier frequencies are near 300 MHz. In this situation, because of dimensional limitations and diminution of the signal magnetic field inside the body, one cannot shift array antenna to higher distance from the body and must keep the antenna array – subject distance as short as possible. Consequently, the gap $d$ between extremely dense array elements can be as small as 3 cm i.e. $d < \lambda/30$ in terms of the operating wavelength $\lambda$. As a result, there will be an overcritical coupling between array elements resulting in the inter-channel scattering and shimming procedure disturbance in absence of the decoupling system.

The presence of the human body beneath the antenna array and the camera wall above it do not allow engineers to decouple such antennas in a straightforward way – inserting an extended absorbing screen in the gap. In the case $d \geq \lambda/10$ one could use the electromagnetic band-gap (EBG) structures for decoupling [8]. However, no one of known EBG structures can decouple two antennas separated by a gap $d = \lambda/30$ because 3 unit cells of the EBG structure is the minimal amount of unit cells required for decoupling [9] and the ultimately miniaturized unit cells have namely the size $\lambda/30$ [9]. In some special (e.g. military) applications, one may use a system of active circuitry for decoupling [10]. However, this technique decreases the power efficiency and requires high-cost maintenance, the criteria which are not applicable for commercially available MRI systems. Passive decoupling techniques suitable for arrays of loop antennas [3-5] are not applicable for decoupling the parallel dipole arrays [6-9]. In [11] authors introduced an elegant method for decoupling two closely located antennas based on parasitic element – a scatterer loaded at its center by a complex load (perhaps with negative real part) that is located exactly in the middle between the antennas. Based on this work, in our papers [12] and [13], we studied theoretically and experimentally the decoupling of two resonant dipoles by a shortcut dipole [12] and by another passive scatterer [13] we called the split-loop resonator (SLR). The operational bandwidth in the case of SLR turned out to be twofold of that corresponding to the decoupling by a dipole, whereas the level of maximal decoupling is the same. However, we proved this advantage of the SLR only in the case when the array of two dipole antennas is located in free space (in our experiment it was located on a low-loss substrate with a low refractive index).

In this paper, we numerically and experimentally investigate the scheme of similar passive decoupling in the case when the antennas are located at a very small height over the phantom having the averaged material properties of a human body. This problem formulation grants a significant novelty to the study. In absence of the phantom, the decoupling scatterer is located exactly in the middle between two antennas. In presence of the phantom, its correct location is a priori not clear since the theory of [11] is not applicable. Moreover, it is not clear a priori will the idea of decoupling by a single scatterer work or not. If the body increases the coupling of two closely located antennas, the single passive scatterer is useless since the only way to increase its own coupling with the antennas is to load it by an impedance with a negative real part that implies a very different technique [11]. If the body beneath the antennas decreases their coupling, the decoupling by a single scatterer is feasible. In this case, we need to decrease the mutual interaction of this scatterer with two antennas shifting it from the antenna plane. However, it is not clear how to shift it: perhaps, closer to the body than the plane of antennas, perhaps farther. We will see that in both cases of a dipole and a SLR used as decoupling scatterers the needed regime holds when we raise the scatterer: both the level of decoupling and the operational bandwidth are sufficient. Furthermore, we observe that the magnetic field, created by the decoupling dipole distorts the antenna field lower than one created by the SLR. Therefore, contrary to [13], the resonant dipole turns out to be advantageous for decoupling compared to the SLR. We confirm our theoretical study by thorough measurements.

---

[*] Corresponding author: Masoud Sharifian Mazraeh Mollaei (Masoud.2.sharifianmazraehmollaei@aalto.fi)
[1] Aalto University, FI-00076, Espoo, Finland. [2] ITMO University, 297101, St. Petersburg, Russia.

## 2. INTERPLAY OF THE PHANTOM AND SCATTERER

In works [12, 13] we reported analytical, numerical and experimental proof of the efficient decoupling for two extremely coupled ($d < \lambda/30$) dipole antennas. This decoupling is granted by a passive scatterer (e.g. a dipole or a SLR) located exactly in the middle between two antennas. In presence of the phantom, the absence of this symmetry can be interpreted in terms of the quasi-static images. Since the height of the dipole antennas $h_d$ over the phantom is electrically very small the interaction of the dipole antennas and the scatterer with the phantom is mainly the near-field one, and the quasi-static image principle is qualitatively applicable.

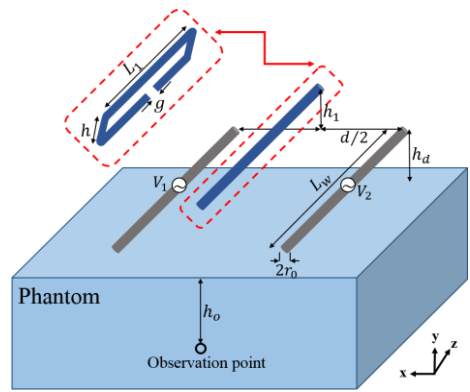

**Figure 1.** Schematic view of active dipoles decoupled by a passive dipole and by a passive SLR in the presence of the phantom.

### 2.1. Impact of the phantom on the decoupling regime

In accordance to this principle, the dipole antennas replaces by a pair of the real dipole and its image located at the depth $h_d$ smaller than $d$. In the image dipole, the current direction inverts and its amplitude is smaller than that in the real dipole (if the phantom permittivity tends to infinity the image dipole amplitude is evidently the same since this case is equivalent to a perfect conductor). Such a pair of dipoles has a nonzero electric dipole moment and a significant magnetic dipole moment. Will two such electric-and magnetic pairs interact weaker or stronger that two parallel (real) dipoles? It is easy to show that in the case of the PEC body they will interact weaker because the electric dipole moment of the pair disappears, and the interaction of two effective magnetic dipoles is lower than that of two resonant electric dipoles. This is so because $h_d$ is very small, and the magnetic dipole creates the near fields lower than those of the electric dipole. In the case of the body phantom, the situation is not very different because the complex permittivity of the human body is very high (of the order of 100). Consequently, the mutual interaction of two antennas shown in Fig. 1 decreases and the required location of the scatterer employed for decoupling must be beyond the plane of two antennas. If we locate the scatterer beneath this plane, it will produce a high parasitic field in the phantom and its interference with the field of the active antennas will be overcritical. Therefore, we have to raise the scatterer over the plane of antennas by a certain height $h_1$. The structure under study with the raised decoupling scatterer is shown in Fig. 1. Since there is no analytical solution for a dipole antenna over a finite phantom, we find the optimal height $h_1$ and other parameters from full-wave numerical simulations which are, further, validated experimentally.

Even in this geometry, the presence of the scatterer results in a certain distortion of the signal magnetic field distribution inside the phantom and may be accompanied by a smaller penetration depth of the signal. In the following, we study the impact of the scatterer on the magnetic field in the phantom.

Next, the impact of the phantom should manifest in the decoupling frequency band and in the matching conditions for antennas in the decoupled regime. It is clear, that the presence of a highly refractive phantom in the near-field zone of the antennas and the scatterer may redshift the resonance frequency (of both dipole antennas and scatterer). Moreover, the power flux into a highly refractive and lossy body should increase the radiation resistance of the antennas that may result in a broader bandwidth for both matching and decoupling of two antennas. We will see below that the increase of the radiation resistance granted by the phantom (sometimes called the Purcell factor [14]) is significant. Therefore, the advantage of the SLR as a decoupling scatterer for a dipole array located in free space – twofold broader operation band compared to the case of decoupling by a dipole – disappears in the presence of the phantom. Due to high radiation resistance, the decoupling band in the matched regime turns out to be sufficient for the normal operation of the antenna array for both dipole scatterer and SLR. A truly important issue is only the parasitic magnetic field of the scatterer in the phantom.

### 2.2. Impact of the decoupling scatterers on distributed magnetic field inside the phantom

For MRI application, the desired decoupling between antennas 1 and 2 means isolating these antenna from each other, while the magnetic field created by antenna 1 inside the phantom in the ideal case remains the same as if antenna 2 was absent. Having in mind this goal, although adding the passive scatterer between dipole antennas 1 and 2 decouples them from each other without a great damage for the operational bandwidth, this scatterer obviously produces a parasitic magnetic field inside the phantom. When the currents in the antenna array in the parallel transmit regime vary in amplitudes and phases so that to better image the target area of the body this parasitic field will change unpredictably. Thus, in the steady regime the secondary radiation of the decoupling scatterer represents a random distortion of the MRI signal field. Therefore, in the present case the best decoupling corresponds to the minimal magnetic field inside the phantom created by the decoupling scatterer. Fortunately, the decoupling scatterer should be located higher over the body than the antennas. Therefore, its parasitic magnetic field inside the phantom is not as high as that of the active antennas. However, it is still comparable and the negative impact is significant. So, we have to choose either a dipole or a SLR namely based on the comparison of their parasitic magnetic fields in the phantom. In the following, we investigate this issue numerically and experimentally.

## 3. NUMERICAL INVESTIGATION

In order to confirm quantitatively all these expectations and to find the optimal value for $h_1$ the structure shown in figure 1 (with either a dipole or a SLR, the gap between the dipole antennas $d = 3$ cm and the height of antennas over the phantom $h_d = 2$ cm) has been simulated using CST Microwave Studio, Time Domain solver. In the simulation, the dipoles and the SLR performed of copper wires have all the same geometric parameters as in works [12, 13]. Active dipoles are excited through lumped ports at their centers. The parameters of the phantom are defined based on the mean human body tissue and correspond to salty water with $\sigma = 1.59$ S/m and $\varepsilon = 78$. Geometric parameters of the antennas, scatterers and phantom are gathered in Table. As a reference structure, we simulated two resonant dipole antenna over the phantom in the absence of the passive scatterer. In the matched case (using virtual matching circuitry in toolbox schematic CST), the transmission coefficient between two antennas achieves -4 dB at the operation frequency and shows a very high coupling between the antennas. After adding the passive scatterers, we performed the simulation of the S-parameters gradually increasing $h_1$. Based on these simulations, optimal $h_1$ for the cases of the dipole scatterer and the SLR are 20 mm and 10 mm, respectively. The simulation results for these optimal cases are shown in figure 2. Similar to [12] and [13], in order to prove the true decoupling, i.e. the decoupling that does not depend on the lumped impedance connected to the antenna input, we plotted the S-parameters of the structure in both matched and mismatched regime of the antennas.

**Table.** Values of the geometric parameters.

| Parameter | Value (mm) | Parameter | Value (mm) |
|---|---|---|---|
| $L_W$ | 500 | $h_P$ | 360 |
| $L_1$ | 290 | $h_o$ | 50 |
| $L_P$ | 400 | $W_p$ | 600 |
| $h$ | 7 | $d$ | 30 |
| $h_d$ | 20 | $g$ | 20 |

Figure 2 completely confirms our previously discussed expectations. We obtain the red shift of resonant frequencies (from 293.2 MHz in [12] to 292 MHz for the case of the decoupling dipole, and from 312.8 MHz in [13] to 291 MHz for the case of the decoupling SLR). We also obtain the wider decoupling band and wider matching band whose intersection gives the operational band of the antenna system. Contrary to decoupling in the absence of the phantom, the decoupling bandwidth granted by the passive SLR is narrower than that one offered by the passive dipole. However, taking into account the needed operational bandwidth of the array for ultra-high field MRI (0.3 MHz) the impact of the phantom makes possible to use both the dipole and SLR for decoupling. The wider operation band offered by the decoupling dipole in the present application is, therefore, not important.

After this numerical investigation, the distribution of the magnetic field inside the phantom has been studied. Since the goal is to approach the magnetic field H in the phantom in presence of antenna 2 and the scatterer to that created by antennas 1 and 2 with equivalent unit current in each, whose magnetic field in the phantom we call the reference distribution of H-field.

To characterize the change of the magnetic field in presence of the scatterer and antenna 2 quantitatively, we define an observation point on the depth $h_o = 50$ mm in the phantom. According to the simulation result, magnitude of H-field for the reference case at this point is equal to 0.23 A/m. Then we simulated the structure in the presence of antenna 2 and the scatterer (either dipole or SLR, located at the corresponding optimal positions for decoupling) and found how the absolute value of H-field changed at the observation point.

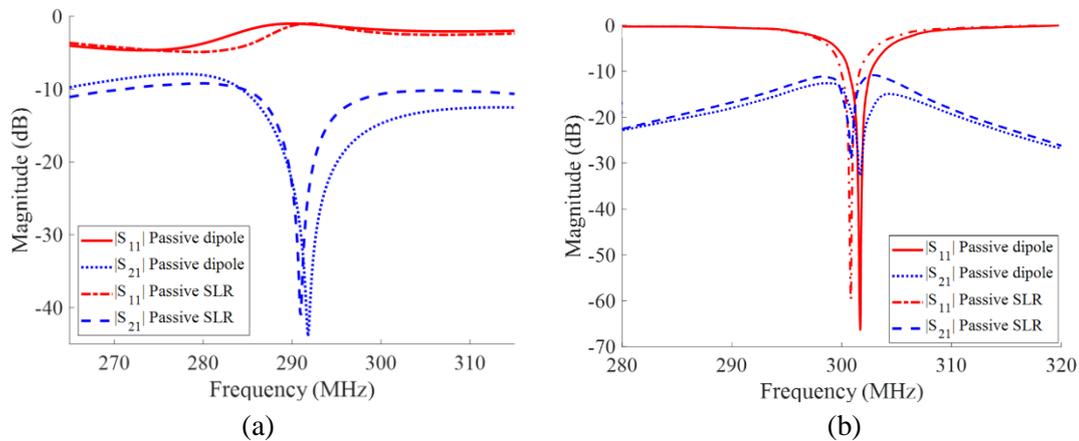

(a)

(b)

**Figure 2.** Simulation results of the decoupled structure by passive dipole and passive SLR with their optimal $h_1$ in the presence of the phantom: **(a)** mismatched regime, **(b)** matched regime. In the matched regime, the bandwidth of S12 granted by the decoupling dipole bandwidth is twofold of that granted by the SLR.

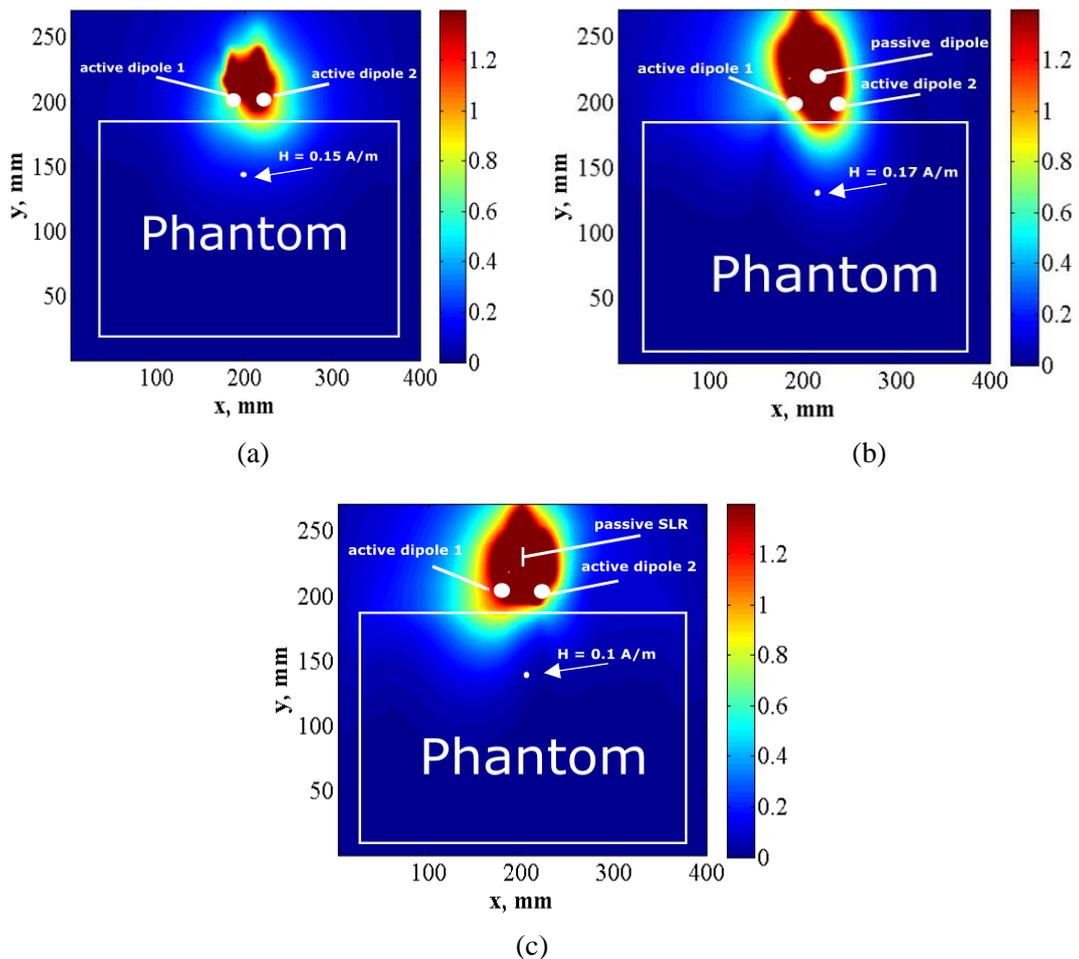

(a)

(b)

(c)

**Figure 3.** H-field distribution inside the phantom: **(a)** reference structure, **(b)** using the passive dipole for decoupling, **(c)** using the passive SLR for decoupling. The value of the signal at the selected depth is marked. In the case **(b)** it is nearly equal to that in the case **(a)**, whereas in the case **(c)** it is lower.

Figure 3 shows simulation results of these structures. We see that the absolute value of the magnetic field at the observation point in the case of the SLR decoupling is lower than the value calculated for the reference structure than the value obtained in the case of the dipole decoupling. The higher parasitic radiation of the SLR into the phantom holds due to lower height of this scatterer needed for decoupling compared to the height of the decoupling dipole.

## 4. EXPERIMENTAL INVESTIGATION

In order to validate these theoretical predictions experimentally, two setups have been fabricated and the operational characteristics measured. The first step of this experiment was preparation and characterization of the phantom with sizes 29 cm ×34 cm and material properties $\sigma = 1.59$ S/m, $\varepsilon = 78$. The phantom contained 0.9% solution of NaCl in distilled water. Complex permittivity of the solution was measured using the precision dielectric measurement system SPEAG DAK-12 and a vector network analyzer Agilent E8362C [15].

The main experimental setup is shown in Fig. 4. It comprises a Vector Network Analyzer (VNA Rohde and Schwarz ZVB-20), two copper dipole antennas excited at their centers through the coaxial cables, the fabricated phantom, and the passive scatterer (either dipole or SLR) supported by foam over the phantom. Geometric parameters of the structure are the same as in Table.

S-parameters of the dipoles (while either the passive dipole or SLR is present) have been measured for different positions of the scatterer over the antenna plane. In a remarkable accordance to our numerical investigation, the true decoupling (both matched and mismatched regimes) holds when the passive dipole is distant by 20 mm and the SLR is distant by 10 mm from the antenna plane. The measurement results for the decoupled structures in the mismatched case are shown in Fig. 5(a) in comparison with simulations. The agreements between simulation and measurement results is very good. For the matched case, the S-parameters of the decoupled structure are shown in Fig. 5(b). Similar to the simulation results, measurement results prove the wider decoupling bandwidth by using the passive SLR (2% relative decoupled bandwidth by adding the passive SLR and 1.4% relative decoupling bandwidth by adding the passive dipole).

Figure 6 pictures the scanning setup. The scanning near-field magnetic probe was a small loop antenna on a movable arm inserted into the phantom. In this measurement, antenna 1, antenna 2 and the probe were connected to channel 1 of VNA, to the matching load and to channel 2 of VNA, respectively. For scanning, the probe moved in a vertical plane orthogonal to the plane of antennas measuring horizontal component of H-field distributed in the cross section of the phantom. Similar to the simulation, first, H-field of the reference structure (equivalently fed antennas 1 and 2) was measured. Further, the H-fields of the decoupled structures (first, by the passive dipole and, second, by the SLR) were measured, respectively. The results are shown in Fig. 7. In the case of the SLR the interference pattern is more spread than in the case of the dipole, that matches the simulations shown in figure 3. though this advantage is not as spectacular as it was in simulations.

Moreover, the penetration depth in our measurements suffers more drastically when the decoupling dipole is replaced by the decoupling SLR. Observed $H$ at the depth of $h_o = 50$ mm for the reference structure, for the decoupled structure by the passive dipole, and for the decoupled structure by the passive SLR are equal to 0.008 A/m, 0.008 A/m and 0.004 A/m, respectively.

With these results in mind, one can conclude that compared to the decoupling dipole, the decoupling SLR results not only in narrower operational band (that is not important in the present application) but also in the higher negative effect on the distributed magnetic field inside the phantom. Both penetration depth and the pattern of the magnetic field are noticeably worse in the case of the decoupling SLR. Notice, that the decoupling dipole does not reduce the penetration depth at all.

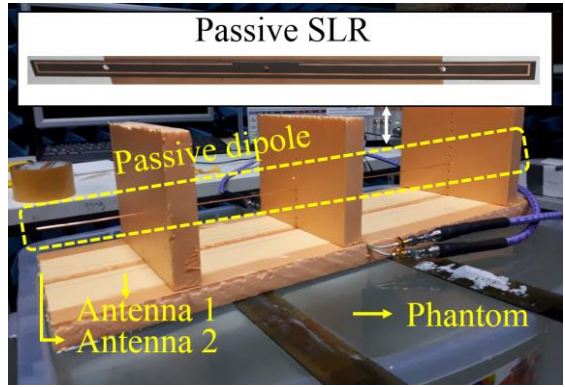

**Figure 4.** Measurement setup for verifying possibility of decoupling in the presence of the phantom.

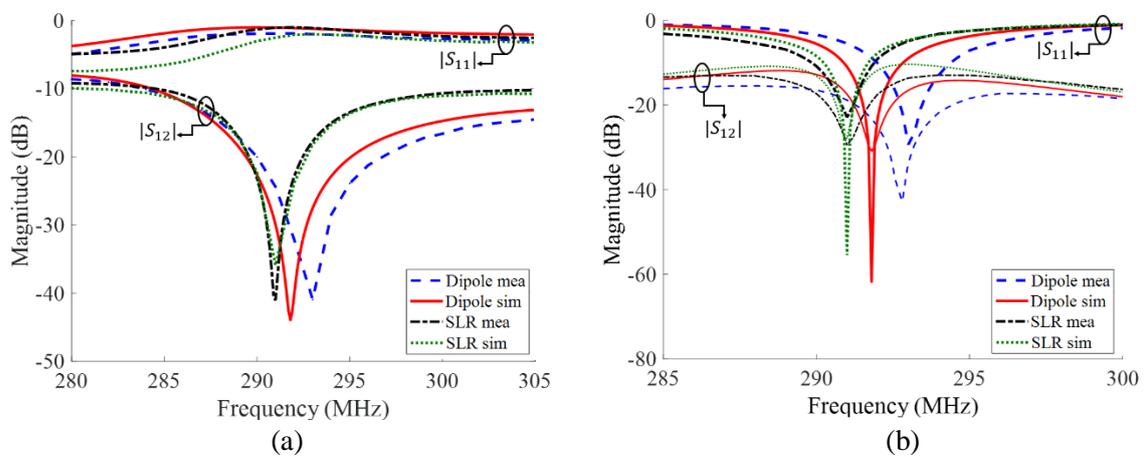

**Figure 5.** Measured and simulated results for the decoupled structure with a passive dipole at height $h_1$=20 mm and a passive SLR at $h_1$=10 mm; **(a)** mismatched regime, **(b)** matched regime.

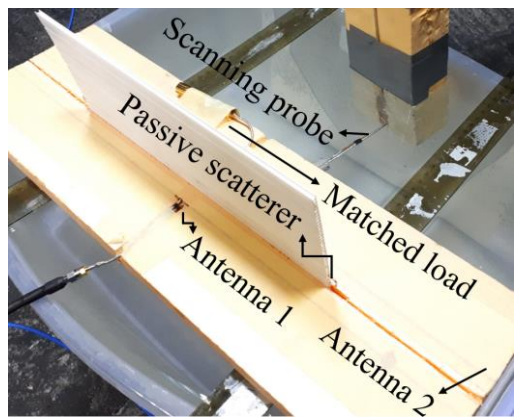

**Figure 6.** Measurement setup for measuring distributed magnetic field inside the phantom.

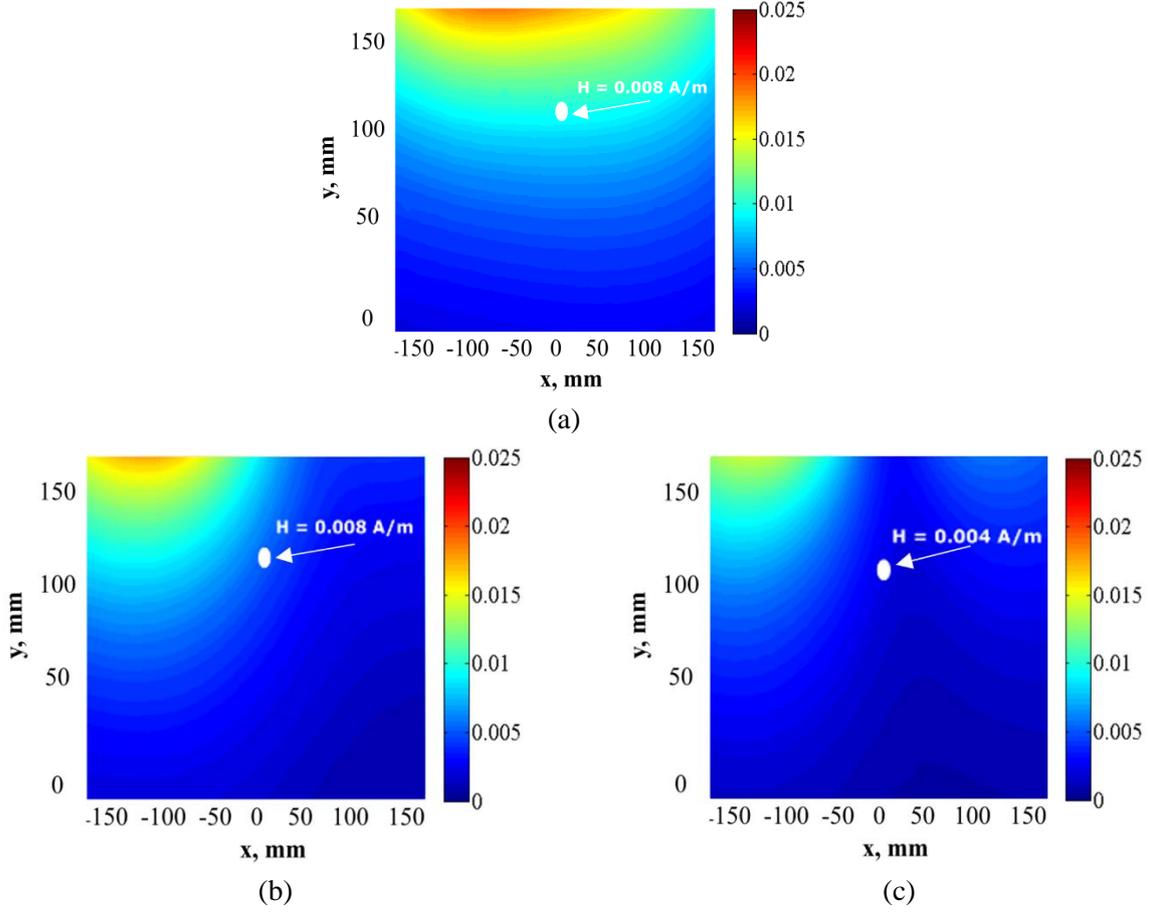

**Figure 7.** Measurement results of the distributed magnetic field inside the phantom; **(a)** the reference structure with two active dipole antennas. **(b)** the structure decoupled by the passive dipole, **(c)** the structure decoupled by the SLR.

## 5. CONCLUSION

We have proved in this work that the passive electromagnetic decoupling of two very closely located (the gap $d$ is below 1/30 of the wavelength) active dipoles by a single scatterer keeps possible in the presence of the human body phantom located very closely ($h_d<d$) to the antennas. Such the situation is typical for a parallel transmit array of dipoles considered as promising counterpart of radiofrequency coils for ultra-high field MRI. We have proved that the decoupling is possible using either by a passive dipole or by a passive SLR located at a small height over the antenna plane. The decoupling holds in both matched and mismatched regime and is, therefore, not sensitive to the lumped impedance connected to the antennas inputs that is the feature of the true decoupling [11]. We have compared the performance of the dipole and the SLR in order to choose the better of two decoupling scatterers. For an array located in free space, the advantageous decoupling can be determined from the consideration of the decoupling level and bandwidth. Here, the presence of the phantom results in the decrease of the coupling accompanied by an increase of the radiation resistance and broadening of the operation band. Therefore, the issue of the operation band becomes not important for comparison and we pay the main attention to the magnetic field inside the phantom, distorted by the decoupling scatterer. From this point of view, the passive dipole turned out to be more advantageous than the SLR. Contrary to [13], decoupling in the dipole array for ultra-high field MRI is preferable utilizing the dipole scatterer rather than the SLR. The replacement of the SLR by the dipole is fully resulting from the presence of the phantom: the decoupling holds for smaller height of the SLR than for the dipole. Therefore, the SLR distorts the signal field more significantly than the dipole.

**ACKNOWLEDGMENT**

Experiments were supported by the Russian Science Foundation (Project No. 18-19-00482). Theoretical studies were supported by the European Union's Horizon 2020 research and innovation program under grant agreement No 736937.

**REFERENCES**

1. Mao, W., Smith, M. B., and Collins, C. M., "Exploring the limits of RF shimming for high-field MRI of the human head," *Magn. Reson. Med.* Vol. 56, No. 4, 918-922, 2006.
2. Ibrahim, T. S. and Tang, L., "Insight into RF power requirements and B1 field homogeneity for human MRI via rigorous FDTD approach," *J. Magn. Reson. Imaging.* Vol. 25, No. 6, 1235-1247, 2007.
3. Avdievich, N. I., Pan, J. W. and Hetherington, H. P., "Resonant inductive decoupling (RID) for transceiver arrays to compensate for both reactive and resistive components of the mutual impedance," *NMR Biomed.* Vol. 26, No. 11, 1547-1554, 2013.
4. Von Morze, C., Tropp, J., Banerjee, S., Xu, D., Karpodinis, K., Carvajal, L., Hess, C. P., Mukherjee, P., Majumdar, S., Vigneron, D. B, "An eight-channel, nonoverlapping phased array coil with capacitive decoupling for parallel MRI at 3 T," *Concepts in Magnetic Resonance B: Magnetic Resonance Engineering,* Vol. 31, 37-43, 2007.
5. Van de Moortele, P. F., Vaughan, T., and Ugurbil, K. A., "A 32-channel lattice transmission line array for parallel transmit and receive MRI at 7 Tesla," *Magn. Reson. Med.,* Vol. 63, No. 6, 1478-1485, 2010.
6. Adriany, G., Van de Moortele, P. F., Wiesinger, F., Moeller, S., Strupp, J. P., Andersen, P., Snyder, C., Zhang, X., Chen, X., Pruessmann, K. P., Boesiger, P., Vaughan, J. T., and Ugurbil, K., "Transmit and receive transmission line arrays for 7 Tesla parallel imaging," *Magn. Reson. Med.,* Vol. 53, 434-445, 2005.
7. Padormo, F., Beqiri, A., Hajnal, J. V., Malik, S. J., "Parallel transmission for ultrahigh-field imaging," *NMR Biomed.* Vol.29, No. 9, 1145-1161, 2015.
8. Hurshkainen, A. A., Derzhavskaya, T. A., Glybovski, S. G., Voogt, I. J., Melchakova, I. V., Van den Berg, C. A. T., and Raaijmakers, A. J. E., "Element decoupling of 7 T dipole body arrays by EBG metasurface structures: Experimental verification," *J. Mag. Reson.* Vol. 269, 87-96, 2016.
9. Li, Q., *Miniaturized DGS and EBG structures for decoupling multiple antennas on compact wireless terminals*, PhD dissertation, Dept. Elect. Eng., Loughborough University, 2012.
10. Fenn, A. J., *Adaptive Antennas and Phased Arrays for Radar and Communications*, Artech House, NY, 2008.
11. Lau, B. K. and Andersen, J. B., "Simple and efficient decoupling of compact arrays with parasitic scatterers," *IEEE Trans. Antennas Propag.,* Vol. 60, No. 11, 464-472, 2012.
12. Mollaei, M. S. M., Hurshkainen, A., Kurdjumov, S., Glybovski, S., and Simovski, C., "Passive electromagnetic decoupling in an active metasurface of dipoles," *Phot. Nanost. Fund. Appl.,* DOI.org/10.1016/j.photonics.2018.10.001.
13. Mollaei, M. S. M., Hurshkainen, A., Glybovski, S., and Simovski, C., "Decoupling of two closely located dipole antennas by a split-loop resonator," *submitted to Radio Sci.,* available in arxiv:1803.00753.

14. Krasnok, A. E., Slobozhanyuk, A. P., Simovski, C. R., Tretyakov, S. A., A.N. Poddubny, A.E. Miroshnichenko, Y.S. Kivshar, and P.A. Belov, "An antenna model for the Purcell effect," *Scientific Reports,* Vol. 5, 1-12, 2015.
15. Giovanetti, G., Frija. F., Menichetti, L., Hartwig, V., Viti, V., and Landini, L., "An Efficient Method for Electrical Conductivity Measurement in the RF Range," *Concepts in Magnetic Resonance Research B*, Vol. 37B, 160-165, 2010.